# Fokker - Planck equation in curvilinear coordinates. Part 2

*Igor A. Tanski*
*Moscow, Russia*
povorot2@infoline.su

V&T

*ABSTRACT*

The aim of this paper is to derive Fokker - Planck equation in curvilinear coordinates using physical argumentation. We get the same result, as in our previous article [1], but for broader class of arbitrary holonomic mechanical systems.

**1. Lagrange equation**

In our previous article [1] we derived Fokker - Planck equation in curvilinear coordinates from its Cartesian form using transformation of space coordinates and velocities. This way does not show physical sense of final equation and involved variables.

We begin from Lagrange equation [2] for holonomic mechanical system

$$\frac{d}{dt}\left(\frac{\partial T}{\partial \dot{q}^i}\right) - \frac{\partial T}{\partial q^i} = Q_i. \tag{1}$$

where $T$ - kinetic energy of the system;
$q^i$ - generalized coordinates;
$Q_i$ - generalized force.

In the following, we use the Einstein summation notation for implicit sums.

We have following expression for kinetic energy

$$T = \frac{1}{2} g_{ij} \dot{q}^i \dot{q}^j. \tag{2}$$

Inertia tensor $g_{ij}$ plays here the same part as the metric tensor in our article [1]. It depends on generalized coordinates and transforms as covariant tensor by transformations of generalized coordinates $q^i$. We use this tensor to lower indices, that is to build one to one correspondence between contravariant and covariant tensors. If the considered mechanical system is a particle of unit mass, both tensors coincide. We therefore use in present article the same notation $g_{ij}$ for inertia tensor as in article [1] for metric tensor.

Velocities $\dot{q}^i$ transform as components of contravariant vector.

Momentum vector

$$p_i = \frac{\partial T}{\partial \dot{q}^i} = g_{ij} \dot{q}^j. \qquad (3)$$

transforms as covariant vector and generalized force $Q_i$ also.

The determinant $g = \det(g_{ij})$ defines 2 measures of volume: measure of volume in configurations space $\sqrt{g} dq^1 dq^2 \cdots$ and measure of volume in velocities space $\sqrt{g} dv^1 dv^2 \cdots$. Both measures are invariants of arbitrary transformations of generalized coordinates.

The product of these measures is measure of volume in phase space. According to Liouville theorem it is invariant of infinitesimal transformation, induced by motion of conservative system.

Let us represent generalized force as a sum of random force $R_i$ and friction force, which is proportional to velocity

$$Q_i = R_i - \alpha g_{ij} \dot{q}^j. \qquad (4)$$

Lagrange equation (1) takes the form

$$\frac{d}{dt}\left(g_{ij} \dot{q}^j\right) - \frac{1}{2} \frac{\partial g_{pq}}{\partial q_i} \dot{q}^p \dot{q}^q = R_i - \alpha g_{ij} \dot{q}^j. \qquad (5)$$

This differential equation of the second order we can write down as a system of two differential equations of the first order. We can do it in many ways, depending on our choice of unknown variable. 2 ways are considered below: contravariant and covariant components of velocities.

## 2. Contravariant components of velocities

Equation (5) with contravariant components of velocities as unknown variables is:

$$\frac{d}{dt}\left(g_{ij} v^j\right) - \frac{1}{2} \frac{\partial g_{pq}}{\partial q_i} v^p v^q = R_i - \alpha g_{ij} v^j. \qquad (6)$$

Expand the first term in (6)

$$\frac{d}{dt}\left(g_{ij} v^j\right) = \frac{dg_{ij}}{dt} v^j + g_{ij} \frac{dv^j}{dt} = \frac{\partial g_{ij}}{\partial q^p} v^p v^j + g_{ij} \frac{dv^j}{dt}, \qquad (7)$$

solve (7) for $\frac{dv^i}{dt}$ and get

$$\frac{dv^j}{dt} = g^{ij}\left(R_i - \alpha g_{is} v^s + \frac{1}{2}\frac{\partial g_{pq}}{\partial q^i} v^p v^q - \frac{\partial g_{is}}{\partial q^p} v^p v^s\right) \tag{8}$$

where $g^{ij}$ is inverse matrix

$$g_{mn} g^{mp} = \delta_n^p, \tag{9}$$

and $\delta_n^p$ is Cronecker delta. It is known [3], that $g^{ij}$ builds contravariant tensor.

Terms with derivatives of inertia tensor are equal to Christoffel symbol [3]

$$\frac{1}{2}\frac{\partial g_{pq}}{\partial q^i} v^p v^q - \frac{\partial g_{is}}{\partial q^p} v^p v^s = \frac{1}{2}\frac{\partial g_{pq}}{\partial q^i} v^p v^q - \frac{1}{2}\frac{\partial g_{qi}}{\partial q^p} v^p v^q - \frac{1}{2}\frac{\partial g_{ip}}{\partial q^q} v^p v^q = -\Gamma_{i,pq} v^p v^q. \tag{10}$$

In this way we obtain the full system with contravariant components of velocities as unknown variables:

$$\frac{dq^i}{dt} = v^i, \tag{11}$$

$$\frac{dv^j}{dt} = g^{ij} R_i - \alpha v^j - \Gamma_{pq}^j v^p v^q. \tag{12}$$

If generalized force is zero, equations (11-12) state, that absolute derivative [3] of $v^j$ vector with respect to metric $g_{ij}$ is zero $\frac{\delta v^j}{\delta t} = 0$. This implies, that the paths of free movement of mechanical system are geodetics of metric $g_{ij}$, defined by its inertia tensor.

Let us consider now not a single system, but ensemble - a large set of noninteracting identical systems. Let us denote $\hat{n}$ - density of systems in joint space of configuration coordinates and velocities. This means, that $\hat{n} dq^1 dq^2 \cdots dv^1 dv^2$ is total number of systems in elementary cell. $\hat{n}$ satisfies evolutionary equation.

$$\frac{\partial \hat{n}}{\partial t} + \frac{\partial(\hat{n} v^i)}{\partial q^i} + \frac{\partial}{\partial v^j}\left[\hat{n}\left(g^{ij} R_i - \alpha v^j - \Gamma_{pq}^j v^p v^q\right)\right] = 0. \tag{13}$$

or

$$\frac{\partial \hat{n}}{\partial t} + v^i \frac{\partial \hat{n}}{\partial q^i} + \frac{\partial}{\partial v^j}\left(\hat{n} g^{ij} R_i\right) - \frac{\partial \hat{n}}{\partial v^j}\left(\alpha v^j + \Gamma_{pq}^j v^p v^q\right) - 3\alpha\hat{n} - 2\hat{n}\Gamma_{jp}^j v^p = 0. \tag{14}$$

We calculate random force $R_i$ from the law

$$n R_i = -k \frac{\partial n}{\partial v^i}. \tag{15}$$

analogous to diffusion (Fick) and heat transfer (Fourier) laws.

This phenomenological law we consider as definition of perfectly random force.

$$\frac{\partial \hat{n}}{\partial t} + v^k \frac{\partial \hat{n}}{\partial q^k} - \Gamma_{pq}^k v^p v^q \frac{\partial \hat{n}}{\partial v^k} - \alpha\, v^k \frac{\partial \hat{n}}{\partial v^k} - 3\,\alpha\,\hat{n} - 2\hat{n}\Gamma_{jp}^j v^p = k\, g^{lk} \frac{\partial^2 \hat{n}}{\partial v^l \partial v^k}\,. \qquad (16)$$

We see, that in the case of zero forces ($\alpha = 0$ and $k = 0$) substitution $\hat{n} = g$ satisfies equation (16). This follows from the known identity [3]

$$\Gamma_{jp}^j = \frac{1}{\sqrt{g}} \frac{\partial \sqrt{g}}{\partial q^p}\,. \qquad (17)$$

This means, that flow without external forces preserves phase space volume. This is Liouville theorem for our case (we do not consider potential forces).

These considerations suggest another definition of density. Let us define

$$\hat{n} = n \det(g_{ij}) = ng. \qquad (18)$$

We get for new density $n$ equation

$$\frac{\partial n}{\partial t} + v^k \frac{\partial n}{\partial x^k} - \Gamma_{pq}^k v^p v^q \frac{\partial n}{\partial v^k} - \alpha\, v^k \frac{\partial n}{\partial v^k} - 3\,\alpha\,n = k\, g^{lk} \frac{\partial^2 n}{\partial v^l \partial v^k}\,. \qquad (19)$$

which coincides with equation (19) from our article [1].

Scaling (18) means, that new density $n$ relates not to product of differentials, but to invariant product of volumes.

## 3. Covariant components of velocities

Covariant components of velocities are also components of momentum vector (see (3)). Therefore we use notation $p_i$ for them.

Equation (5) takes the form

$$\frac{dp_i}{dt} - \frac{1}{2} \frac{\partial g_{pq}}{\partial q_i} \dot{q}^p \dot{q}^q = R_i - \alpha p_i. \qquad (20)$$

This leads to following system of differential equations:

$$\frac{dp_i}{dt} = \frac{1}{2} \frac{\partial g_{pq}}{\partial q_i} g^{pm} g^{qn} p_m p_n + R_i - \alpha p_i, \qquad (21)$$

$$\frac{dq^i}{dt} = g^{im} p_m. \qquad (22)$$

We use the ensemble concept again. Let us denote $n$ - density of systems in joint space of configuration coordinates and momentum vectors. This means, that $n \, dq^1 dq^2 \cdots dp_1 dp_2 \cdots$ is total number of systems in elementary cell.

The measure of volume in velocities space is $\frac{1}{\sqrt{g}} dp_1 dp_2 \cdots$, therefore the measure of volume in phase space is $dq^1 dq^2 \cdots dp_1 dp_2$ and $n$ so defined coincides with density from the section 2.

$n$ satisfies following evolutionary equation:

$$\frac{\partial n}{\partial t} + \frac{\partial (n g^{im} p_m)}{\partial q^i} + \frac{\partial}{\partial p_i}\left[ n \left( \frac{1}{2} \frac{\partial g_{pq}}{\partial q_i} g^{pm} g^{qn} p_m p_n + R_i - \alpha p_i \right) \right] = 0. \qquad (23)$$

Second term in (23) expands as follows:

$$\frac{\partial (n g^{im} p_m)}{\partial q^i} = g^{im} p_m \frac{\partial n}{\partial q^i} + n p_m \frac{\partial g^{im}}{\partial q^i}. \qquad (24)$$

Third term in (23) simplifies to

$$\frac{\partial}{\partial p_i}\left( \frac{1}{2} \frac{\partial g_{pq}}{\partial q_i} g^{pm} g^{qn} p_m p_n \right) = \frac{1}{2} \frac{\partial g_{pq}}{\partial q_j} g^{pj} g^{qn} p_n + \frac{1}{2} \frac{\partial g_{pq}}{\partial q_i} g^{pm} g^{qi} p_m = \frac{\partial g_{pq}}{\partial q_j} g^{pj} g^{qn} p_n. \qquad (25)$$

So we have

$$\frac{\partial n}{\partial t} + g^{im} p_m \frac{\partial n}{\partial q^i} + n p_m \frac{\partial g^{im}}{\partial q^i} + \frac{\partial n}{\partial p_i}\left( \frac{1}{2} \frac{\partial g_{pq}}{\partial q_i} g^{pm} g^{qn} p_m p_n - \alpha p_i \right) + \qquad (26)$$

$$+ n \frac{\partial g_{pq}}{\partial q_j} g^{pj} g^{qn} p_n - 3\alpha n + \frac{\partial}{\partial p_i}(n R_i) = 0.$$

The sum of coefficients by $n$ is equal to zero

$$\frac{\partial g^{in}}{\partial q^i} + \frac{\partial g_{pq}}{\partial q_j} g^{pj} g^{qn} = 0. \qquad (27)$$

To prove this, we begin from identity $g^{in} g_{nk} = \delta^i_k$. Differentiate this by $q^i$ and obtain

$$\frac{\partial g^{in}}{\partial q^i} g_{nk} + g^{in} \frac{\partial g_{nk}}{\partial q^i} = 0. \qquad (28)$$

Then lift the index and get the desired result

$$\frac{\partial g^{in}}{\partial q^i} + g^{kp} g^{in} \frac{\partial g_{nk}}{\partial q^i} = 0. \tag{29}$$

For the random force we have expression (15). It follows

$$nR_i = -k \frac{\partial n}{\partial v^i} = -k \frac{\partial n}{\partial p_j} \frac{\partial p_j}{\partial v^i} = -k g_{ij} \frac{\partial n}{\partial p_j}. \tag{30}$$

So we have

$$\frac{\partial n}{\partial t} + g^{im} p_m \frac{\partial n}{\partial q^i} + \frac{\partial n}{\partial p_i} \left( \frac{1}{2} \frac{\partial g_{pq}}{\partial q_i} g^{pm} g^{qn} p_m p_n \right) - \alpha p_i \frac{\partial n}{\partial p_i} - 3\alpha n = k g_{ij} \frac{\partial^2 n}{\partial p_i \partial p_j}. \tag{31}$$

Now it remains only one step to the final result:

$$\frac{\partial n}{\partial t} + g^{mk} p_m \frac{\partial n}{\partial q^k} + \Gamma^q_{kl} g^{pl} p_p p_q \frac{\partial n}{\partial p_k} - \alpha\, p_k \frac{\partial n}{\partial p_k} - 3\,\alpha\, n = k\, g_{lk} \frac{\partial^2 n}{\partial p_l \partial p_k}. \tag{32}$$

To prove equivalence of (31) and (32), we only need to check identity

$$p_m p_n \left( \frac{1}{2} \frac{\partial g_{pq}}{\partial q_i} g^{pm} g^{qn} - \Gamma^m_{is} g^{sn} \right) = 0. \tag{33}$$

It follows directly from definition of Christoffel symbol

$$\Gamma^m_{is} = \frac{1}{2} \left( \frac{\partial g_{ki}}{\partial q_s} + \frac{\partial g_{sk}}{\partial q_i} - \frac{\partial g_{is}}{\partial q_k} \right) g^{km}. \tag{34}$$

This gives

$$-p_m p_n \frac{1}{2} \left( \frac{\partial g_{ki}}{\partial q_s} - \frac{\partial g_{is}}{\partial q_k} \right) g^{km} g^{sn} = 0. \tag{35}$$

But expression inside of parenthesis is antisymmetric on indices $k, s$ and expression outside of parenthesis is symmetric. Therefore the sum is zero, q.e.d.

**DISCUSSION**

We proved, that equations (19) and (32) are forms of Fokker - Planck equation for contravariant and covariant components of velocities respectively. They coincide with forms, derived in our article [1].

We proved (19) and (32) not only for particle movement in curvilinear coordinates, but for broader class of arbitrary holonomic mechanical systems.

Usually authors [4, 5] begin their studies from Hamilton's equations. This is not our way, because we wish to consider friction force and random force from the very beginning. Therefore we need instrument like Lagrange equations.

Our equations (19) and (32) are analogous to Liouville equation [5] for density in phase space. This builds connection of Fokker - Planck equation approach with another works in the field of statistical mechanics .

______________________________